\documentstyle[11pt]{article}
\begin{document}

\centerline{C. Y. Chen}
\centerline{Dept. of Physics, Beijing University of Aeronautics}
\centerline{and Astronautics, Beijing 100083, PRC}
\vskip 20pt
\centerline{Email: cychen@buaa.edu.cn}

\vfill
\noindent {\bf Abstract:} 
A nonperturbative procedure of solving the time-dependent Schr\"odinger
equation, called the multi-projection approach or phase dynamics of
quantum mechanics, is derived and illustrated. In addition to introducing 
a method with that time-dependent systems become solvable (under the 
assumption that corresponding time-independent systems are solvable), the
new approach unveils several misconcepts related to the usual wavefunction 
expansion and the standard perturbation theory. 

\newpage

\section{Introduction}
The standard methodology of quantum dynamics, initially due to 
Dirac\cite{dirac}, assumes that a wave function 
can always be expanded into a series of eigenfunctions 
\begin{equation}\label{series}
\sum C_k(t) e^{-i\omega_k t}W_k({\bf r}) ,\end{equation}
where $W_k({\bf r})$ represent a set of eigenfunctions, which is,
according 
to the customary understanding, associated with the initial Hamiltonian 
$H(t_0)$ of the quantum system. By inserting the expansion into the 
time-dependent Schr\"odinger equation and doing some mathematical 
manipulations, 
a set of coupled differential equations is obtained as 
\begin{equation}\label{ordi}
 i\hbar\frac{dC_k}{dt}= \sum_m C_m e^{i(\omega_k-\omega_m)t} V_{km}, 
\end{equation}
where $V_{km}$ are the matrix elements of the Hamiltonian
variation $V(t)\equiv H(t)-H(t_0)$. A common concept, taught 
everywhere and established
solidly in our mind, is that the expansion (\ref{series}) is the formal
solution of the wave function and the equation set (\ref{ordi})
describes the system's behavior as exactly as the original Schr\"odinger
equation does. Many perturbative and nonperturbative approaches are based 
on or related to this concept. (As an example, see the derivation of the 
Liouville theorem in quantum statistics.)

Our studies, however, showed that
the concept outlined above, while enjoying tremendous successes, involved
serious difficulties\cite{chen}\cite{chen1}. 
Some of them are the following. 
\begin{itemize}
\item  In numerical computations, the sum of squared moduli of all  
coefficients, {\it i.e.}, $\sum |C_k(t)|^2$, can easily go to infinity 
if $V(t)$ is not truly small. That is to say, the Dirac theory, like
many other types of perturbation theories, suffers from the divergence 
difficulty.
\item The notion associated with (\ref{series}) and (\ref{ordi}) asserts 
that if a quantum system, say a harmonic oscillator, is initially in the
ground state, it will make transition, partially though, to a higher 
state at the next moment, 
and then to a higher higher state at the next next 
moment. The dynamical process will never terminate if $H(t)-H(t_0)$ is 
nonzero as $t\rightarrow \infty$ (another 
symptom of the divergence). In contrast with that, 
a general analysis of quantum mechanics
states that whenever the Hamiltonian $H(t)$ 
becomes steady, equal to $H(t_0)$ or not, the system will instantly be 
settled in a stationary state.
\item According to the state-transition picture provided by the Dirac 
theory, a ground-state system will never lose its energy under any 
circumstances. This conclusion does not appear to be in harmony with 
other theoretical and experimental observations.
\item The fact that an electromagnetic field can be represented by 
different gauge fields implies that there are an infinite number of 
equation sets taking the form (\ref{ordi}) for one definite disturbance. 
Direct numerical calculations show that these equation sets are not 
equivalent to each other. 
\end{itemize}

Many questions then arise, of which some must be of fundamental interest
to the community. How does the Dirac formalism, derived mathematically 
from the Schr\"odinger equation, involve so many difficulties? Why can 
the theory, running into problems ultimately, still offer good results 
under certain conditions? What on earth are these conditions?
Is there a method by that the time-dependent Schr\"odinger equation can 
be solved more adequately?
Among them, the last question seems to be the most essential one. 
If we find out a proper way of solving this fundamental equation, 
many questions, in particular those related to the Dirac perturbation
theory, can be answered accordingly.   
                                                                      
Although a different perturbation theory\cite{chen1}, which corresponded
to a similar
theory in classical mechanics\cite{chen2} and suffered less problems,
was put forward, the issue was far from
cleared-up and the aforementioned questions remained to be open.

In this paper we wish to report on a nonperturbative procedure of 
solving the 
time-dependent Schr\"odinger equation. Interestingly, the new procedure, 
called the multi-projection approach or 
phase dynamics of quantum mechanics, is still based on Dirac's 
idea: solving a nonstationary quantum system with knowledge of 
stationary quantum systems. But, unlike its predecessor, 
this approach suffers from
no divergence difficulty and exhibits surprising effectiveness. Any 
dynamical system, subject to weak or strong fields, becomes analyzable
and calculable provided that the corresponding 
stationary states can be solved analytically or numerically.

The structure of this paper is the following. Section 2 introduces our
new approach, which is based on an assumption that the 
time-dependent Hamiltonian can be approximated by its stepwise 
varying counterparts. In Sec. 3, similarities and dissimilarities of 
the proposed approach to the influential path-integral approach are 
remarked. Simple applications are presented in Section 4, where 
effectiveness of the new approach is illustrated. Section 
5 rederives the Dirac perturbation theory and answers several related
questions. Sec. 6 gives discussions on the common concept of 
wavefunction expansion. Sec. 7 concludes the paper.

\section{Multi-projection approach}
In this section, we try to determine the wave function $\Psi(t)$ on the 
condition that the initial wave function $\Psi(t_0)$ and  the 
time-dependent Hamiltonian $H(t)$ are given. 

According to the basic formalism of 
quantum mechanics, the solution of the Schr\"odinger equation can be 
written as 
\begin{equation}\label{solu1} \Psi(t) = e^{-\frac i\hbar \int_{t_0}^t 
H(\tau)d\tau} \Psi(t_0). \end{equation}
The solution above is just a formal one except in the situation where
the involved Hamiltonian is independent of time. 
To generally evaluate it, 
we slice the time interval from $t_0$ to $t$ into $n$ segments, equal
to each other or not, as 
\begin{equation}\label{inter} \Delta t_1=t_1-t_0,\;\; 
\cdots,\;\;\cdots,\;\;\Delta t_n=t-t_{n-1}. \end{equation}
The formal solution (\ref{solu1}) can then be rewritten as 
\begin{equation}\label{solu2} \Psi(t) = e^{-\frac i\hbar \int_{t_{n-1}}^t 
H(\tau)d\tau} \cdots\cdots e^{-\frac i\hbar \int_{t_0}^{t_1} H(\tau)d\tau} 
\Psi(t_0). \end{equation}
Or, in terms of the intermediate quantum states,
\begin{equation}\label{solu3} 
\Psi(t_j)= e^{-\frac i\hbar 
\int_{t_{j-1}}^{t_j} H(\tau)d\tau} \Psi(t_{j-1})\quad (j=1,2,\cdots).
\end{equation}
Referring to Fig. 1, we explore the possibility to replace the 
Hamiltonian $H(t)$ with its stepwise varying approximation $\hat H(t)$.
Without losing generality, consider a charged particle in a 
time-dependent electromagnetic field, whose Hamiltonian reads
\begin{equation} H(t)=\frac 1{2m}({\bf p}-\frac Qc{\bf A})^2+\Phi.
\end{equation}
 It is rather obvious that if
$\Phi\not= 0$ and $\bf A=0$, the replacement of $H(t)$ with $\hat 
H(t)$ is justified by the observation that the two Hamiltonians represent 
roughly the same physical system. For the case in that both the scalar 
potential $\Phi$ and the vector potential $\bf A$ are nonzero, the 
situation becomes somewhat complicated. It is well-known, the vector 
potential $\bf A$ can be separated into two parts: the longitudinal field 
and the transverse field\cite{jck}. By realizing that the longitudinal  
vector field can cause trouble to our formalism 
(the reason for saying it will be clear), 
only the scalar field and transverse vector 
field will be treated herein. This should be allowable in 
view of that an 
appropriate gauge transformation can always make the longitudinal vector 
field vanish.
 Under these understandings, we define the Hamiltonian 
$\hat H(t)$ as 
\begin{equation} \label{defh} {\hat H}_j(t)=\frac 1{\Delta t_j}
\int^{t_j}_{t_{j-1}} H(t) dt \quad {\rm for}\;\; (t_{j-1}<t<t_j);
\end{equation}
or, even in a simpler way, ${\hat H}_j(t)=[H(t_{j-1})+H(t_j)]/2$. 
Thus, during each of the time segments defined by (\ref{inter}) the new 
Hamiltonian ${\hat H}_j$ is independent of time and the typical 
intermediate state in (\ref{solu3}) becomes
\begin{equation}\label{solu4} 
\Psi(t_j)\approx e^{-\frac i\hbar {\hat H}_j \Delta t_j} \Psi(t_{j-1}). 
   \end{equation} 
The problem is reduced to a familiar one related to solving stationary 
systems. For each of the intermediate Hilbert space associated with 
${\hat H}_j$, we have normalized intermediate energy eigenfunctions 
\begin{equation} {\hat H}_j W_k^j({\bf r})= E_k^j W_k^j({\bf 
r}). \end{equation}
After the wave function $\Psi(t_{j-1})$ is known, the wave function 
$\Psi(t_j)$ can be expressed by 
\begin{equation} \label{solu6} \Psi(t_j) = \sum C^j_k e^{-i \omega_k^j 
\Delta t_j } W_k^j({\bf r}),\end{equation}
where $\omega_k^j=E_k^j/\hbar$ and $C^j_k$ is determined by a projection 
\begin{equation} \label{solu5} C^j_k= \int \Psi(t_{j-1}) W^{j*}_k({\bf 
r}) d{\bf r}.\end{equation}
Or, in the Dirac notation,
\begin{equation}\label{dirac}
|t_j\rangle = \sum_{k_j} \langle k_j|t_{j-1}\rangle e^{-i\omega_k^j
 \Delta t_j} |k_j \rangle,\end{equation}
where $|k_j\rangle$ stands for $W_k^j$. 
Eq. (\ref{dirac}) shows that for a short time interval 
a dynamical system and its correponding stationary system evolve
in the same way. 
By repeating the step as presented above, the wave function at the time 
$t$ can be expressed as
\begin{equation}\label{final}
|t\rangle = \sum_{k_n}\cdots \sum_{k_1}  e^{-i\sum_j \omega_k^j 
\Delta t_j } | k_n \rangle \langle k_n| k_{n-1}\rangle 
\cdots \langle k_1| t_0 \rangle. \end{equation}
For convenience of discussion, we will name $| k_n \rangle \langle 
k_n| k_{n-1}\rangle \cdots \langle k_1| t_0 \rangle$ as a 
minute component of $|t_0 \rangle$, by which we refer to the fact 
that such a component is in usual situations very small.
It is easy to prove that the sum of all the minute 
components, with no phase factors involved, is just 
equal to the initial wavefunction.

In deriving the above formalism, it has been assumed that the wave 
function $\Psi(t<t_0)$ does not contain any phase factor in the form 
$e^{if(t,{\bf r})}$ where $f$ depends on $t$ and ${\bf r}$ explicitly. 
Generally speaking, $e^{if(t,{\bf r})}$ is not a uniformly continuous 
function with respect to $t$ and ${\bf r}$ and great difficulties will
be encountered if we try to expand a wavefunction having such  
phase factor.
A related and important point is that since we have assumed that all 
$\hat H_j$ contain no longitudinal vector fields, all the intermediate 
wave functions are also free from such phase factors. 

Numerically speaking, the accuracy of this method largely depends on the
step size. If all steps in a calculation approach infinitesimal ones, the 
resultant wavefunction will approach the real dynamical one.  

For a dynamical system, the energy at any moment can 
naturally be defined by the intermediate 
Hamiltonian and the intermediate wavefunction. In the sense of taking
limit, this definition is an accurate one.
                                     
Before finishing this section, we wish to point out that 
this approach offers a different picture about quantum 
dynamics. The Dirac theory leads one to imagine how a 
nonstationary system makes transition from one eigenstate to another. 
In contrast with that, the multi-projection approach 
reveals that a nonstationary system
is subjected to no other change than that 
each minute component of the wavefunction acquires its own
dynamical phase factor. 
To be consistent with the spirit of  
enlightening discussions about Berry's phase\cite{berry}, the obtained 
formalism may be called phase dynamics of quantum mechanics.

\section{Equivalence to the path-integral approach}
In this section, the path-integral approach\cite{path}\cite{shan} is 
briefly reviewed and then the relation between the path-integral approach
and the multi-projection approach is remarked.

In the path-integral approach we have
\begin{equation} \label{sim1}
\Psi(x,t)=\int \langle x,t|x_0,t_0\rangle \Psi(x_0,t_0) dx_0,
\end{equation}
where $\langle x,t|x_0,t_0\rangle$ is named as the propagator. 
(For simplicity, only the one-dimensional case is considered.)
It is found that the propagator takes the form
\begin{equation} 
\langle x,t|x_0,t_0\rangle =\int^x_{x_0} e^{iS[x(\tau)]/\hbar}
D[x(\tau)],\end{equation}
where $D[x(\tau)]$ is the measure associated with all possible
paths $x(\tau)$ and the action $S$ is the integral of the Lagrangian 
along the path 
\begin{equation} S=\int^t_{t_0} L(\tau)d\tau=\int^t_{t_0}\left[\frac 12
 mv^2 -V(x(\tau),\tau)\right]d\tau.\end{equation}
The formalism above is quite formal unless the discrete form of it is 
written out. By slicing the time $(t-t_0)$ into $N$ equal segments and
denoting $(t-t_0)/N$ by $\varepsilon$, the action integral becomes 
\begin{equation} \label{pro1}
S=\sum\limits_{j=0}^{N-1}\left[ \frac {m(x_{j+1}-x_j)^2}
{2\varepsilon} -\varepsilon V(x^\prime,\varepsilon^\prime)\right]
\end{equation}
where $x^\prime$ and $\varepsilon^\prime$ take values within the 
intervals $(x_j,x_{j+1})$ and $(t_j,t_{j+1})$ respectively. 
Upon this, the propogator becomes
\begin{equation} 
\lim\limits_{N\rightarrow\infty}A\int\limits_{-\infty}^{+\infty}
\cdot\cdot \int\limits_{-\infty}^{+\infty} 
\exp \sum\limits_{j=0}^{N-1}\frac i\hbar\left[ \frac {m(x_{j+1}-x_j)^2}
{2\varepsilon} -\varepsilon V(x^\prime,\varepsilon^\prime)\right]
dx_1\cdot \cdot dx_{N-1}, \end{equation} 
where the factor $A$ can be determined by the limit 
$\langle x,\varepsilon|x_0,0\rangle\rightarrow \delta(x-x_0)$.  

For purposes of this paper, we wish to take a look at how the 
path-integral 
formalism yields the standard Schr\"odinger equation\cite{shan}.
Consider the propagator associated with one slice of time after $t=0$ 
\begin{equation}\label{pro2}
\langle x,\varepsilon|x_0,0\rangle= \left(\frac m{2\pi\hbar i\varepsilon}
\right)^{1/2}\exp\left\{ \frac i\hbar \left[\frac{m(x-x_0)^2} 
{2\varepsilon} -\varepsilon V(x^\prime, \varepsilon^\prime)
\right]\right\}, \end{equation}
which means
\begin{eqnarray}\label{pro3} 
&& \hspace{-20pt}
\Psi(x,\varepsilon)=\displaystyle \left(\frac m{2\pi\hbar i\varepsilon}
 \right)^{1/2} \nonumber  \\
&& \quad\times
\int^\infty_{-\infty} \exp\left\{ \frac i\hbar \left[\frac{m(x-x_0)^2} 
{2\varepsilon} -\varepsilon V(x^\prime, \varepsilon^\prime)\right]\right\}
\Psi(x_0,0)dx_0. \end{eqnarray}
By noting that the region $(x-x_0)^2\sim 2\varepsilon \hbar \pi/m$ 
gives the main contribution to the integral, we obtain, after taking
approximations,
\begin{equation} \label{simf}
i\hbar \frac{\Psi(x,\varepsilon)-\Psi(x,0)}\varepsilon
\approx \left[-\frac{\hbar^2}{2m}\frac{\partial^2}{\partial x^2} 
+V(x,\varepsilon^\prime)\right]\Psi(x,0), 
\end{equation}
which is just the finite differential form of the Schr\"odinger equation.

The similarities between the path-integral and  the
multi-projection are found by the following observations. (i) 
Both approaches acquire clear meaning in their discrete forms. (ii) Both 
approaches are approximate theories in their discrete forms and convergent
to the exact theory as steps of the discrete forms become infinitesimal. 
(iii) For a short time interval, both approaches show indifference to 
whether or not the quantum system is truly time-dependent.   
In the path-integral approach outlined in this section, if we replace
the Lagragian with its stepwise varying approximation, all the formulas,
in particular (\ref{pro2}), (\ref{pro3}) and (\ref{simf}), make no 
essential change;
as the steps go smaller and smaller, exactly the same limits will be 
obtained. All these imply that  
the equivalence of these two approaches is indeed there.

It is now worth mentioning dissimilarities.
The path-integral approach formulates the time unitary transformation of 
wave function in the ordinary spatial space, 
whereas the multi-projection approach does the same job between a series 
of Hilbert spaces whose bases are formed by eigenfunctions of the 
intermediate Hamiltonians. Due to
this difference, the formalism in the path-integral approach takes a form
of probability integral, while the formalism in the 
multi-projection approach takes a form of a series of definite projections. 
In terms of studying practical systems, it is expected that the two 
provide different computational ease.

Finally, a note about gauge. It is rather necessary for both the 
approaches to assume no longitudinal vector field to exist. 

\section{Simple applications}
Some concrete examples, in which effectiveness 
of the multi-projection approach manifests itself, are presented in
this section.  

Firstly, consider a case in that the potential of a 
harmonic oscillator is subject to a ``sudden'' change and the Hamiltonian  
reads  
\begin{equation}\label{hamil} H(t)=\frac {p^2}{2m} +\frac {S(t)}2 k x^2 
,\end{equation}
with 
\begin{equation} \label{st}
 S(t)=\left\{\begin{array}{ll}  1 & (t\leq 0)\\
      \eta &(t> 0), \end{array}\right.   \end{equation}
in which $\eta$ is a positive constant other than $1$. Physical intuition 
says that, if $\eta>1$ the total energy of the system must go larger 
(the spatial room for the oscillator is ``compressed''); 
if $\eta<1$ the total energy must go smaller. 

Though the situation given above 
is terribly simple, it still defines a dynamical system that needs 
to be solved in one way or another. 
As indicated in the introduction, if the standard method is applied
serious difficulties arise. In what follows we 
try to use our multi-projection approach. 

Assume that the system is initially in the ground state, 
namely\cite{schiff}
\begin{equation}\label{q1}
\Psi(0)= N_0(\alpha)\exp(-\alpha^2 x^2/2),\end{equation}
and the wave function after $t=0$ takes the form
\begin{equation} \label{q2}
\Psi(t)=\sum_n C_n e^{-i(n+\frac 12 )\omega^\prime t} 
N_n(\alpha^\prime)H_n(\alpha^\prime x) e^{-\frac 12 {\alpha^\prime}^2 
x^2}.\end{equation} 
The notation in Eqs. (\ref{q1}) and (\ref{q2}) is rather standard:
 $\alpha= (mk/\hbar^2)^{1/4}$, $\alpha^\prime = \sqrt[4]\eta \alpha $, 
$\omega^\prime = \sqrt\eta \omega=\sqrt{\eta k/m}$ and
$$  N_n(\alpha)=\left(\frac \alpha{\sqrt\pi 2^n n!}\right)^{\frac 12},
\quad H_{n+1}(\xi)=2\xi H_n-H_{n-1}\;({\rm with}\; H_0=1).$$
The energy of the system at $t=0$ is known to be $E_0=0.5\hbar \omega$.
By making use of (\ref{solu5}), we obtain, for $\eta=0.5^2$, 
\begin{equation} C_0=\left(\frac 89\right)^{\frac 14},\quad C_1=0,\quad 
C_2=-\frac 23 \left(\frac 1{72}\right)^{\frac 14},\quad C_3=0\cdots.
\end{equation}
Numerically, we have 
\begin{equation} C_0\approx 0.9710,\; C_2\approx -0.2289, \; C_4\approx 
0.0661 ,\; C_6\approx -0.0201; \end{equation}
these imply $\sum |C_n|^2\approx 1 $ and the final energy becomes
\begin{equation}
\langle E \rangle=\sum |C_n|^2(n+0.5)\hbar \omega^\prime 
\approx  0.6246 E_0.\end{equation}
Similarly, we obtain, for $\eta=0.9^2$ and $\eta=1.1^2$ respectively,
\begin{equation}
 \langle E \rangle \approx 0.9050 E_0 \quad{\rm and}\quad
 \langle E \rangle \approx 1.1050 E_0.
\end{equation} 
It is now obvious that with the multi-projection approach 
(i) the normalization condition for wave function, namely  
$\sum|C_n|^2=1$, is automatically satisfied; (ii) whenever the Hamiltonian
becomes steady the system is settled in a stationary state; (iii) 
even if a quantum system is initially in a ground state, the total 
energy of it can decrease or increase; (iv) no gauge issues pose 
problems if we let the longitudinal vector potential vanish always. 
In other words, all the difficulties 
troubling the standard dynamical theory disappear completely. 

The following similar example shows that the proposed method is a good mean
for evaluating the phase shift of a quantum system. If the oscillator 
expressed by (\ref{hamil}) is again in the ground state at $t=0$ and 
the function $S(t)$ in (\ref{hamil}) is, instead of (\ref{st}), 
\begin{equation} \label{st1} 
S(t)=\left\{\begin{array}{ll}  1 & (t\leq 0)\\
\eta &(0<t<T)\\ 1&(t\geq T) , \end{array}\right.   \end{equation}
the formulation of the above example can still be employed except that a
projection at $t=T$ onto the original Hilbert space is needed. In general,
the final wave function involves energy transition and phase  
shift. As a special case, we consider the situation in which $T=4\pi
/\omega^\prime$. Simple 
calculation tells us that at $t=T$ the wave function will 
come back to its original value and original phase (by virtue of 
$e^{i2\pi}=1$) as if the system has been completely ``frozen''. 
But, on the other hand, one finds that if the system gets no 
disturbed, it will  during the same time acquire
 the phase factor
\begin{equation} \label{phase}
\exp(- i\omega T/2)=\exp(-i 2\pi /\sqrt{\eta}). \end{equation}
It is then obvious that the disturbance (\ref{st1}) makes 
the wavefunction have the additional 
phase factor $\exp(i 2\pi /\sqrt{\eta})$.

An inspection of the last example states that if a perturbation 
is capable of changing the system's eigenfrequencies, it can in 
general make the wavefunction have 
an additional phase shift, irrespective of whether or not there 
is energy transition. As shown by experiments\cite{tomita}, such 
phase shift has physical effects and should be treated with care. 

\section{Rederivation of the Dirac perturbation theory}
In this section we rederive the Dirac perturbation theory with help of
our proposed approach. The purpose of doing that is two folds. One is 
to illustrate the analytical ability of the approach;
and the other is to answer questions related to the Dirac
perturbation theory.

To obtain an analytical formalism comparable
to the Dirac one, we presuppose that (i) the quantum system of interest
is initially in the $n$th eigenstate of $H_0$, the perturbation applies
at $t=0$ and vanishes completely at $t=T$;
(ii) the perturbation can be approximated by a series of pulses whose
values rise sharply and vanish sharply, as illustrated in Fig. 2 
(the pulse-like
perturbation has the same physical effects as the real perturbation);
(iii) the system's wavefunction involves no additional 
phase-factor shift before and after each of the pulses. With these  
assumptions adopted, the derivation here can be deemed as a simple 
application of the ``sudden approximation'' due to Pauli\cite{pauli}.

Consider one specific pulse of the perturbation that exists between 
$t^\prime$ and $t^\prime+\Delta t^\prime$. 
The leading term of the wave function at $t=t^\prime$ is 
$e^{-i\omega_n t^\prime}|n\rangle$, at $t=T$ it becomes, 
by repeated use of (\ref{dirac}),
\begin{eqnarray}\label{arrow}
&&\hspace{-10pt}
 e^{-i\omega_n t^\prime}|n\rangle \rightarrow \displaystyle\sum_k 
e^{-i\omega_n t^\prime} 
e^{-i\omega_k \Delta t^\prime}\langle k |n\rangle |k\rangle \nonumber\\
&& \quad \rightarrow \displaystyle\sum_{k,m} e^{-i\omega_n t^\prime} 
e^{-i\omega_k \Delta 
t^\prime} e^{-i\omega_m (T-t^\prime-\Delta t^\prime)} \langle m |k\rangle  
\langle k |n\rangle |m\rangle, \end{eqnarray}
where $|m\rangle$ stand for eigenfunctions defined by the 
Hamiltonian $H_0$ and $|k\rangle$ eigenfunctions defined by the 
intermediate Hamiltonian within the pulse. Note that in (\ref{arrow}) 
all the phase factors take on their unperturbed values before $t^\prime$
and after $t^\prime+\Delta t^\prime$ as if no other pulses exist, 
which should in general be 
regarded as a rough approximation, as revealed in the last section.
Since $\Delta t^\prime$ is 
short, the following approximation is acceptable
\begin{equation}
\sum_k \langle m |k\rangle e^{-i(\omega_k-\omega_m)\Delta t^\prime} 
\langle k| n\rangle\approx \left\langle m\left|1-\frac i\hbar [H(t^\prime)
-H_0]\Delta t^\prime\right| n\right\rangle, \end{equation}
thus, the $m$th coefficient of the wave function at $t=T$ is 
\begin{equation} \label{bm0}
 b_m = -\frac i\hbar \Delta t^\prime \langle m| V|n\rangle 
e^{-i(\omega_n-\omega_m) t^\prime} e^{-i\omega_m T},\quad (m\not= n)
\end{equation}
where $V \equiv H(t^\prime)-H_0$. Taking contributions from all pulses 
into account, we obtain at $t=T$
\begin{equation}\label{bm}
b_m=-\frac i\hbar e^{-i\omega_m T} 
\int^T_0 V_{mn} e^{-i(\omega_n-\omega_m) t^\prime} dt^\prime
\quad (m\not= n) .\end{equation}
Except the phase factor $\exp(-i\omega_m T)$, the formula above is 
consistent with the well-known one. 

Unlike the standard derivation in textbooks, the present derivation
clears up several subtle and important things, of which some were
addressed in the introduction. 
Firstly, it indicates that if the final 
Hamiltonian $H(t>T)$ is not the same as the initial Hamiltonian, 
this formalism will assume that ``perturbation pulses'' are always 
there and thus yield misleading results. 
Secondly, it stresses that if the perturbation, represented by the 
Hamiltonian variation $V(t)$, is not relatively small 
or if the action time of the perturbation is not relatively short, 
the leading term of the wavefunction, including its phase factor, 
will in general be disturbed significantly and 
the accuracy of the formalism will partly or entirely be demolished.  
Thirdly, it suggests that higher-order solutions directly given by the 
Dirac perturbation theory are not meaningful.
Finally, it implies that in order for the formalism to hold, an 
appropriate gauge, like the one defined in Sec. 2, needs to be adopted.

\section{Discussion on the wavefunction expansion}
Let us return to the subject put forth at the very beginning. What have
been presented in this paper, together with those in Ref. 2, actually 
bring out that the usual concept about the wavefunction expansion bears  
problematic aspects. The formulation in Sec. 2 clearly shows   
that after a quantum system leaves its initial state, the  
eigenfunctions and eigenfrequencies associated with the initial 
Hamiltonian become out-of-date. If we 
forcefully use them to express the wavefunction at the later times, the 
coefficients of the would-be series have to adjust 
themselves violently---so violently that they cannot be regarded as
continuous variables in the usual sense. If the aim of a calculation
is to determine these coefficients continuously, the situation 
will soon become out of control: the more terms are taken 
into account, the faster temporal scales and the larger spatial scales 
get involved, the bigger and more serious errors thus enter the 
scheme.

In other words, although one can, at any fixed moment,  
expand a dynamical wavefunction into a series of the initial 
eigenfunctions as, up to a phase factor,
\begin{equation}\label{series11}
\sum C_n(t) W_n({\bf r}) \quad{\rm or}\quad
\sum C_n(t) e^{-i\omega_n t}W_n({\bf r}) ,\end{equation}
it is not appropriate to assume that a dynamical wavefunction
can be represeted by an expansion that takes the form 
(\ref{series11}) and has continuously time-varying coefficients. 

The arguments that have just been presented unveil, in another 
perspective, that the concepts introduced
by this approach---intermediate Hilbert spaces, 
intermediate eigenfrequencies and discrete projections
between intermediate Hilbert spaces---are more fundamental than 
they appear to be. 
(Also, it justifies the path-integral approach, whose
objective is to formulate a wavefunction at certain discrete
moments.) 

\section{Summary}

We have proposed a
nonperturbative procedure, called the multi-projection approach,
which in a way unifies treatments of different 
quantum systems: stationary and nonstationary, weakly-disturbed and 
strongly-disturbed. In view of that
knowledge about stationary systems has been accumulated 
for long and many systems with strong fields need to be studied, 
such unification appears to be quite desirable.

The equivalence of the multi-projection approach to the 
path-integral approach also suggests that the proposed method
may find its applications in a variety of quantum fields, though
it is still too early to say what exactly they are.

This paper has shown that a dynamical process of quantum system is,
rather strikingly, characterized by the phase-factor evolution of 
each minute component of the wavefunction. 
This ``state 
transition'' picture is quite different from the standard one 
provided by the Dirac perturbation theory. The standard picture,
associated usually with one fixed Hilbert space and paying much 
less attention to disturbance of phase factors, is applicable
only if some conditions are simultaneously satisfied, which in 
general include: 
the final Hamiltonian of the system is the same as the initial 
Hamiltonian, the perturbation is relatively small,
and the action time of the perturbation is relatively short.

We have pointed out that a dynamical wavefunction cannot be
represented by a wavefunction expansion that has 
continuously time-varying coefficients.

Stimulating discussion with Professor Han-ying Guo is gratefully 
acknowledged. This paper is partly supported by School of Science, 
BUAA, PRC.

\newpage
\setlength{\unitlength}{0.020in} 
\begin{picture}(220,140)
\put(52,110){\makebox(35,8)[l]{$H(t)$}}
\put(178,16){\makebox(35,8)[l]{$t$}}
\put(174,20.3){\vector(1,0){1}}
\put(50.3,114){\vector(0,1){1}}
\put(45,20){\line(1,0){128}}
\put(50,15){\line(0,1){98}}
\put( 50, 37.499){\line(1,0){6.2}}
\put( 56, 51.744){\line(1,0){6.2}}
\put(56.000, 37.499){\line(0,1){14.245}}
\put( 62, 66.823){\line(1,0){6.2}}
\put(62.000, 51.744){\line(0,1){15.079}}
\put( 68, 80.111){\line(1,0){6.2}}
\put(68.000, 66.823){\line(0,1){13.288}}
\put( 74, 89.301){\line(1,0){6.2}}
\put(74.000, 80.111){\line(0,1){9.190}}
\put( 80, 93.548){\line(1,0){6.2}}
\put(80.000, 89.301){\line(0,1){4.247}}
\put( 86, 93.889){\line(1,0){6.2}}
\put(86.000, 93.548){\line(0,1){0.341}}
\put( 92, 92.728){\line(1,0){6.2}}
\put(92.000, 92.728){\line(0,1){1.161}}
\put( 98, 92.643){\line(1,0){6.2}}
\put(98.000, 92.643){\line(0,1){0.085}}
\put(104, 95.105){\line(1,0){6.2}}
\put(104.000, 92.643){\line(0,1){2.462}}
\put(110, 99.751){\line(1,0){6.2}}
\put(110.000, 95.105){\line(0,1){4.646}}
\put(116, 104.559){\line(1,0){6.2}}
\put(116.000, 99.751){\line(0,1){4.809}}
\put(122, 107.003){\line(1,0){6.2}}
\put(122.000, 104.559){\line(0,1){2.443}}
\put(128, 104.648){\line(1,0){6.2}}
\put(128.000, 104.648){\line(0,1){2.354}}
\put(134, 97.597){\line(1,0){6.2}}
\put(134.000, 97.597){\line(0,1){7.051}}
\put(140, 87.203){\line(1,0){6.2}}
\put(140.000, 87.203){\line(0,1){10.394}}
\put(146, 76.053){\line(1,0){6.2}}
\put(146.000, 76.053){\line(0,1){11.150}}
\put(152, 66.536){\line(1,0){6.2}}
\put(152.000, 66.536){\line(0,1){9.517}}
\put(158,20){\line(0,1){46.53}}

\put(50.000, 32.000){\circle*{1}}
\put(50.436, 32.900){\circle*{1}}
\put(50.873, 33.800){\circle*{1}}
\put(51.299, 34.704){\circle*{1}}
\put(51.725, 35.609){\circle*{1}}
\put(52.140, 36.518){\circle*{1}}
\put(52.556, 37.428){\circle*{1}}
\put(52.971, 38.338){\circle*{1}}
\put(53.374, 39.253){\circle*{1}}
\put(53.777, 40.168){\circle*{1}}
\put(54.171, 41.087){\circle*{1}}
\put(54.564, 42.007){\circle*{1}}
\put(54.958, 42.926){\circle*{1}}
\put(55.342, 43.849){\circle*{1}}
\put(55.726, 44.772){\circle*{1}}
\put(56.104, 45.699){\circle*{1}}
\put(56.481, 46.625){\circle*{1}}
\put(56.859, 47.551){\circle*{1}}
\put(57.230, 48.479){\circle*{1}}
\put(57.602, 49.407){\circle*{1}}
\put(57.973, 50.336){\circle*{1}}
\put(58.339, 51.266){\circle*{1}}
\put(58.706, 52.197){\circle*{1}}
\put(59.070, 53.128){\circle*{1}}
\put(59.434, 54.060){\circle*{1}}
\put(59.798, 54.991){\circle*{1}}
\put(60.161, 55.923){\circle*{1}}
\put(60.523, 56.855){\circle*{1}}
\put(60.886, 57.787){\circle*{1}}
\put(61.249, 58.719){\circle*{1}}
\put(61.611, 59.650){\circle*{1}}
\put(61.974, 60.582){\circle*{1}}
\put(62.339, 61.513){\circle*{1}}
\put(62.704, 62.445){\circle*{1}}
\put(63.072, 63.374){\circle*{1}}
\put(63.440, 64.304){\circle*{1}}
\put(63.808, 65.234){\circle*{1}}
\put(64.181, 66.162){\circle*{1}}
\put(64.555, 67.089){\circle*{1}}
\put(64.928, 68.017){\circle*{1}}
\put(65.309, 68.941){\circle*{1}}
\put(65.691, 69.866){\circle*{1}}
\put(66.080, 70.787){\circle*{1}}
\put(66.470, 71.708){\circle*{1}}
\put(66.860, 72.629){\circle*{1}}
\put(67.262, 73.544){\circle*{1}}
\put(67.664, 74.460){\circle*{1}}
\put(68.079, 75.369){\circle*{1}}
\put(68.494, 76.279){\circle*{1}}
\put(68.910, 77.189){\circle*{1}}
\put(69.343, 78.090){\circle*{1}}
\put(69.777, 78.991){\circle*{1}}
\put(70.230, 79.882){\circle*{1}}
\put(70.684, 80.774){\circle*{1}}
\put(71.159, 81.653){\circle*{1}}
\put(71.634, 82.533){\circle*{1}}
\put(72.135, 83.399){\circle*{1}}
\put(72.636, 84.264){\circle*{1}}
\put(73.167, 85.111){\circle*{1}}
\put(73.698, 85.959){\circle*{1}}
\put(74.265, 86.782){\circle*{1}}
\put(74.832, 87.606){\circle*{1}}
\put(75.443, 88.397){\circle*{1}}
\put(76.093, 89.157){\circle*{1}}
\put(76.742, 89.917){\circle*{1}}
\put(77.443, 90.629){\circle*{1}}
\put(78.191, 91.293){\circle*{1}}
\put(78.939, 91.956){\circle*{1}}
\put(79.752, 92.538){\circle*{1}}
\put(80.616, 93.040){\circle*{1}}
\put(81.522, 93.460){\circle*{1}}
\put(82.465, 93.793){\circle*{1}}
\put(83.435, 94.033){\circle*{1}}
\put(84.423, 94.180){\circle*{1}}
\put(85.421, 94.238){\circle*{1}}
\put(86.420, 94.217){\circle*{1}}
\put(87.416, 94.127){\circle*{1}}
\put(88.405, 93.981){\circle*{1}}
\put(89.388, 93.795){\circle*{1}}
\put(90.365, 93.581){\circle*{1}}
\put(91.338, 93.352){\circle*{1}}
\put(92.311, 93.121){\circle*{1}}
\put(93.286, 92.898){\circle*{1}}
\put(94.265, 92.694){\circle*{1}}
\put(95.249, 92.521){\circle*{1}}
\put(96.240, 92.389){\circle*{1}}
\put(97.237, 92.308){\circle*{1}}
\put(98.236, 92.286){\circle*{1}}
\put(99.235, 92.332){\circle*{1}}
\put(100.228, 92.450){\circle*{1}}
\put(101.208, 92.643){\circle*{1}}
\put(102.172, 92.908){\circle*{1}}
\put(103.115, 93.241){\circle*{1}}
\put(104.032, 93.637){\circle*{1}}
\put(104.950, 94.033){\circle*{1}}
\put(105.827, 94.514){\circle*{1}}
\put(106.673, 95.047){\circle*{1}}
\put(107.494, 95.617){\circle*{1}}
\put(108.296, 96.215){\circle*{1}}
\put(109.081, 96.834){\circle*{1}}
\put(109.866, 97.454){\circle*{1}}
\put(110.631, 98.097){\circle*{1}}
\put(111.385, 98.754){\circle*{1}}
\put(112.134, 99.417){\circle*{1}}
\put(112.883, 100.080){\circle*{1}}
\put(113.628, 100.746){\circle*{1}}
\put(114.377, 101.409){\circle*{1}}
\put(115.133, 102.064){\circle*{1}}
\put(115.889, 102.719){\circle*{1}}
\put(116.660, 103.355){\circle*{1}}
\put(117.449, 103.970){\circle*{1}}
\put(118.259, 104.556){\circle*{1}}
\put(119.093, 105.106){\circle*{1}}
\put(119.928, 105.657){\circle*{1}}
\put(120.804, 106.138){\circle*{1}}
\put(121.717, 106.544){\circle*{1}}
\put(122.664, 106.865){\circle*{1}}
\put(123.638, 107.085){\circle*{1}}
\put(124.632, 107.187){\circle*{1}}
\put(125.630, 107.158){\circle*{1}}
\put(126.616, 106.994){\circle*{1}}
\put(127.570, 106.700){\circle*{1}}
\put(128.482, 106.290){\circle*{1}}
\put(129.343, 105.785){\circle*{1}}
\put(130.155, 105.202){\circle*{1}}
\put(130.919, 104.557){\circle*{1}}
\put(131.683, 103.913){\circle*{1}}
\put(132.388, 103.204){\circle*{1}}
\put(133.049, 102.453){\circle*{1}}
\put(133.710, 101.703){\circle*{1}}
\put(134.326, 100.916){\circle*{1}}
\put(134.942, 100.129){\circle*{1}}
\put(135.517, 99.311){\circle*{1}}
\put(136.069, 98.476){\circle*{1}}
\put(136.620, 97.642){\circle*{1}}
\put(137.149, 96.794){\circle*{1}}
\put(137.678, 95.945){\circle*{1}}
\put(138.187, 95.084){\circle*{1}}
\put(138.696, 94.224){\circle*{1}}
\put(139.190, 93.354){\circle*{1}}
\put(139.683, 92.484){\circle*{1}}
\put(140.164, 91.608){\circle*{1}}
\put(140.645, 90.731){\circle*{1}}
\put(141.118, 89.850){\circle*{1}}
\put(141.590, 88.968){\circle*{1}}
\put(142.057, 88.084){\circle*{1}}
\put(142.523, 87.199){\circle*{1}}
\put(142.990, 86.315){\circle*{1}}
\put(143.452, 85.428){\circle*{1}}
\put(143.914, 84.541){\circle*{1}}
\put(144.376, 83.654){\circle*{1}}
\put(144.837, 82.767){\circle*{1}}
\put(145.301, 81.881){\circle*{1}}
\put(145.765, 80.995){\circle*{1}}
\put(146.233, 80.111){\circle*{1}}
\put(146.701, 79.228){\circle*{1}}
\put(147.176, 78.347){\circle*{1}}
\put(147.650, 77.467){\circle*{1}}
\put(148.133, 76.592){\circle*{1}}
\put(148.616, 75.716){\circle*{1}}
\put(149.110, 74.846){\circle*{1}}
\put(149.604, 73.977){\circle*{1}}
\put(150.110, 73.115){\circle*{1}}
\put(150.617, 72.253){\circle*{1}}
\put(151.140, 71.400){\circle*{1}}
\put(151.662, 70.547){\circle*{1}}
\put(152.202, 69.706){\circle*{1}}
\put(152.743, 68.865){\circle*{1}}
\put(153.304, 68.037){\circle*{1}}
\put(153.866, 67.210){\circle*{1}}
\put(154.451, 66.399){\circle*{1}}
\put(155.056, 65.603){\circle*{1}}
\put(155.661, 64.807){\circle*{1}}
\put(156.291, 64.030){\circle*{1}}
\put(156.920, 63.253){\circle*{1}}
\put(157.579, 62.501){\circle*{1}}
\end{picture}

\vskip 30pt
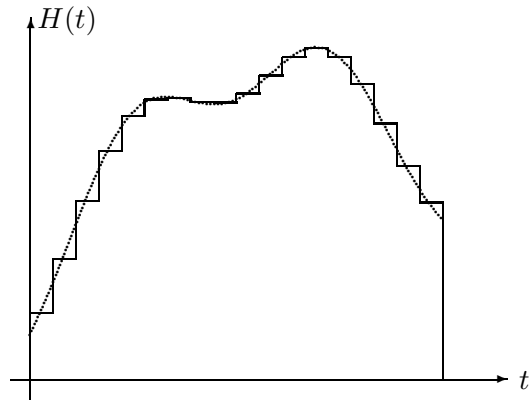
\noindent Fig. 1, A typical time-dependent Hamiltonian and its 
stepwise-varying approximation. 

\setlength{\unitlength}{0.020in} 
\begin{picture}(220,150)
\put(52,110){\makebox(35,8)[l]{$V(t)\equiv H(t)-H_0$}}
\put(178,16){\makebox(35,8)[l]{$t$}}
\put(51,12){\makebox(35,8)[l]{$0$}}
\put(157,13){\makebox(35,8)[l]{$T$}}
\put(174,20.3){\vector(1,0){1}}
\put(50.3,114){\vector(0,1){1}}

\put(45,20){\line(1,0){128}}
\put(50,15){\line(0,1){98}}
\put( 50, 29.423){\line(1,0){5}}
\put( 50, 20){\line(0,1){9.423}}
\put( 55, 20){\line(0,1){9.423}}
\put( 56, 50.220){\line(1,0){5}}
\put( 56, 20){\line(0,1){30.220}}
\put( 61, 20){\line(0,1){30.220}}
\put( 62, 66.235){\line(1,0){5}}
\put( 62, 20){\line(0,1){46.235}}
\put( 67, 20){\line(0,1){46.235}}
\put( 68, 75.935){\line(1,0){5}}
\put( 68, 20){\line(0,1){55.935}}
\put( 73, 20){\line(0,1){55.935}}
\put( 74, 79.094){\line(1,0){5}}
\put( 74, 20){\line(0,1){59.094}}
\put( 79, 20){\line(0,1){59.094}}
\put( 80, 75.875){\line(1,0){5}}
\put( 80, 20){\line(0,1){55.875}}
\put( 85, 20){\line(0,1){55.875}}
\put( 86, 68.678){\line(1,0){5}}
\put( 86, 20){\line(0,1){48.678}}
\put( 91, 20){\line(0,1){48.678}}
\put( 92, 59.713){\line(1,0){5}}
\put( 92, 20){\line(0,1){39.713}}
\put( 97, 20){\line(0,1){39.713}}
\put( 98, 51.468){\line(1,0){5}}
\put( 98, 20){\line(0,1){31.468}}
\put(103, 20){\line(0,1){31.468}}
\put(104, 45.942){\line(1,0){5}}
\put(104, 20){\line(0,1){25.942}}
\put(109, 20){\line(0,1){25.942}}
\put(110, 44.075){\line(1,0){5}}
\put(110, 20){\line(0,1){24.075}}
\put(115, 20){\line(0,1){24.075}}
\put(116, 46.105){\line(1,0){5}}
\put(116, 20){\line(0,1){26.105}}
\put(121, 20){\line(0,1){26.105}}
\put(122, 50.400){\line(1,0){5}}
\put(122, 20){\line(0,1){30.400}}
\put(127, 20){\line(0,1){30.400}}
\put(128, 54.943){\line(1,0){5}}
\put(128, 20){\line(0,1){34.943}}
\put(133, 20){\line(0,1){34.943}}
\put(134, 57.342){\line(1,0){5}}
\put(134, 20){\line(0,1){37.342}}
\put(139, 20){\line(0,1){37.342}}
\put(140, 54.889){\line(1,0){5}}
\put(140, 20){\line(0,1){34.889}}
\put(145, 20){\line(0,1){34.889}}
\put(146, 46.344){\line(1,0){5}}
\put(146, 20){\line(0,1){26.344}}
\put(151, 20){\line(0,1){26.344}}
\put(152, 31.044){\line(1,0){5}}
\put(152, 20){\line(0,1){11.044}}
\put(157, 20){\line(0,1){11.044}}
\put(50.000, 20.000){\circle*{1}}
\put(50.249, 20.968){\circle*{1}}
\put(50.499, 21.937){\circle*{1}}
\put(50.748, 22.905){\circle*{1}}
\put(50.997, 23.874){\circle*{1}}
\put(51.249, 24.841){\circle*{1}}
\put(51.501, 25.809){\circle*{1}}
\put(51.753, 26.777){\circle*{1}}
\put(52.009, 27.744){\circle*{1}}
\put(52.264, 28.711){\circle*{1}}
\put(52.520, 29.677){\circle*{1}}
\put(52.775, 30.644){\circle*{1}}
\put(53.035, 31.610){\circle*{1}}
\put(53.295, 32.575){\circle*{1}}
\put(53.555, 33.541){\circle*{1}}
\put(53.815, 34.507){\circle*{1}}
\put(54.081, 35.471){\circle*{1}}
\put(54.346, 36.435){\circle*{1}}
\put(54.612, 37.399){\circle*{1}}
\put(54.878, 38.363){\circle*{1}}
\put(55.151, 39.325){\circle*{1}}
\put(55.423, 40.287){\circle*{1}}
\put(55.696, 41.249){\circle*{1}}
\put(55.969, 42.211){\circle*{1}}
\put(56.250, 43.171){\circle*{1}}
\put(56.532, 44.130){\circle*{1}}
\put(56.813, 45.090){\circle*{1}}
\put(57.104, 46.047){\circle*{1}}
\put(57.394, 47.004){\circle*{1}}
\put(57.684, 47.961){\circle*{1}}
\put(57.975, 48.917){\circle*{1}}
\put(58.277, 49.871){\circle*{1}}
\put(58.578, 50.824){\circle*{1}}
\put(58.880, 51.777){\circle*{1}}
\put(59.195, 52.727){\circle*{1}}
\put(59.509, 53.676){\circle*{1}}
\put(59.823, 54.625){\circle*{1}}
\put(60.151, 55.570){\circle*{1}}
\put(60.479, 56.515){\circle*{1}}
\put(60.807, 57.459){\circle*{1}}
\put(61.151, 58.398){\circle*{1}}
\put(61.496, 59.337){\circle*{1}}
\put(61.840, 60.276){\circle*{1}}
\put(62.204, 61.207){\circle*{1}}
\put(62.567, 62.139){\circle*{1}}
\put(62.931, 63.070){\circle*{1}}
\put(63.318, 63.992){\circle*{1}}
\put(63.706, 64.914){\circle*{1}}
\put(64.116, 65.826){\circle*{1}}
\put(64.527, 66.738){\circle*{1}}
\put(64.937, 67.649){\circle*{1}}
\put(65.380, 68.546){\circle*{1}}
\put(65.823, 69.442){\circle*{1}}
\put(66.300, 70.321){\circle*{1}}
\put(66.777, 71.200){\circle*{1}}
\put(67.293, 72.057){\circle*{1}}
\put(67.809, 72.913){\circle*{1}}
\put(68.371, 73.740){\circle*{1}}
\put(68.933, 74.567){\circle*{1}}
\put(69.554, 75.350){\circle*{1}}
\put(70.226, 76.090){\circle*{1}}
\put(70.898, 76.830){\circle*{1}}
\put(71.645, 77.494){\circle*{1}}
\put(72.457, 78.076){\circle*{1}}
\put(73.331, 78.559){\circle*{1}}
\put(74.261, 78.923){\circle*{1}}
\put(75.234, 79.145){\circle*{1}}
\put(76.230, 79.209){\circle*{1}}
\put(77.223, 79.109){\circle*{1}}
\put(78.189, 78.857){\circle*{1}}
\put(79.111, 78.475){\circle*{1}}
\put(79.984, 77.989){\circle*{1}}
\put(80.857, 77.503){\circle*{1}}
\put(81.655, 76.901){\circle*{1}}
\put(82.401, 76.236){\circle*{1}}
\put(83.109, 75.529){\circle*{1}}
\put(83.816, 74.823){\circle*{1}}
\put(84.482, 74.077){\circle*{1}}
\put(85.119, 73.306){\circle*{1}}
\put(85.756, 72.536){\circle*{1}}
\put(86.366, 71.743){\circle*{1}}
\put(86.976, 70.951){\circle*{1}}
\put(87.561, 70.140){\circle*{1}}
\put(88.132, 69.319){\circle*{1}}
\put(88.704, 68.499){\circle*{1}}
\put(89.264, 67.670){\circle*{1}}
\put(89.824, 66.842){\circle*{1}}
\put(90.374, 66.007){\circle*{1}}
\put(90.925, 65.172){\circle*{1}}
\put(91.469, 64.333){\circle*{1}}
\put(92.012, 63.493){\circle*{1}}
\put(92.554, 62.653){\circle*{1}}
\put(93.097, 61.813){\circle*{1}}
\put(93.640, 60.974){\circle*{1}}
\put(94.187, 60.136){\circle*{1}}
\put(94.733, 59.299){\circle*{1}}
\put(95.286, 58.465){\circle*{1}}
\put(95.838, 57.632){\circle*{1}}
\put(96.401, 56.805){\circle*{1}}
\put(96.963, 55.978){\circle*{1}}
\put(97.540, 55.161){\circle*{1}}
\put(98.129, 54.353){\circle*{1}}
\put(98.718, 53.545){\circle*{1}}
\put(99.326, 52.751){\circle*{1}}
\put(99.933, 51.957){\circle*{1}}
\put(100.566, 51.183){\circle*{1}}
\put(101.222, 50.428){\circle*{1}}
\put(101.877, 49.673){\circle*{1}}
\put(102.565, 48.947){\circle*{1}}
\put(103.282, 48.250){\circle*{1}}
\put(103.999, 47.553){\circle*{1}}
\put(104.760, 46.905){\circle*{1}}
\put(105.559, 46.304){\circle*{1}}
\put(106.393, 45.753){\circle*{1}}
\put(107.264, 45.262){\circle*{1}}
\put(108.169, 44.839){\circle*{1}}
\put(109.108, 44.495){\circle*{1}}
\put(110.074, 44.240){\circle*{1}}
\put(111.061, 44.083){\circle*{1}}
\put(112.058, 44.029){\circle*{1}}
\put(113.056, 44.079){\circle*{1}}
\put(114.044, 44.229){\circle*{1}}
\put(115.014, 44.471){\circle*{1}}
\put(115.960, 44.794){\circle*{1}}
\put(116.906, 45.117){\circle*{1}}
\put(117.808, 45.546){\circle*{1}}
\put(118.678, 46.039){\circle*{1}}
\put(119.522, 46.575){\circle*{1}}
\put(120.346, 47.142){\circle*{1}}
\put(121.152, 47.733){\circle*{1}}
\put(121.959, 48.324){\circle*{1}}
\put(122.745, 48.942){\circle*{1}}
\put(123.521, 49.573){\circle*{1}}
\put(124.291, 50.211){\circle*{1}}
\put(125.059, 50.852){\circle*{1}}
\put(125.827, 51.492){\circle*{1}}
\put(126.596, 52.131){\circle*{1}}
\put(127.372, 52.761){\circle*{1}}
\put(128.158, 53.380){\circle*{1}}
\put(128.944, 53.999){\circle*{1}}
\put(129.749, 54.591){\circle*{1}}
\put(130.575, 55.155){\circle*{1}}
\put(131.424, 55.683){\circle*{1}}
\put(132.298, 56.167){\circle*{1}}
\put(133.202, 56.596){\circle*{1}}
\put(134.134, 56.955){\circle*{1}}
\put(135.095, 57.229){\circle*{1}}
\put(136.079, 57.401){\circle*{1}}
\put(137.077, 57.452){\circle*{1}}
\put(138.072, 57.370){\circle*{1}}
\put(139.046, 57.150){\circle*{1}}
\put(139.980, 56.799){\circle*{1}}
\put(140.915, 56.447){\circle*{1}}
\put(141.763, 55.920){\circle*{1}}
\put(142.539, 55.290){\circle*{1}}
\put(143.254, 54.592){\circle*{1}}
\put(143.917, 53.844){\circle*{1}}
\put(144.581, 53.096){\circle*{1}}
\put(145.183, 52.299){\circle*{1}}
\put(145.786, 51.501){\circle*{1}}
\put(146.330, 50.662){\circle*{1}}
\put(146.873, 49.823){\circle*{1}}
\put(147.367, 48.954){\circle*{1}}
\put(147.861, 48.085){\circle*{1}}
\put(148.316, 47.194){\circle*{1}}
\put(148.770, 46.303){\circle*{1}}
\put(149.193, 45.397){\circle*{1}}
\put(149.615, 44.491){\circle*{1}}
\put(150.011, 43.573){\circle*{1}}
\put(150.408, 42.655){\circle*{1}}
\put(150.804, 41.737){\circle*{1}}
\put(151.173, 40.807){\circle*{1}}
\put(151.542, 39.878){\circle*{1}}
\put(151.911, 38.949){\circle*{1}}
\put(152.256, 38.010){\circle*{1}}
\put(152.601, 37.071){\circle*{1}}
\put(152.945, 36.133){\circle*{1}}
\put(153.270, 35.187){\circle*{1}}
\put(153.594, 34.241){\circle*{1}}
\put(153.918, 33.295){\circle*{1}}
\put(154.226, 32.343){\circle*{1}}
\put(154.533, 31.392){\circle*{1}}
\put(154.84, 30.440){\circle*{1}}
\put(155.13, 29.484){\circle*{1}}
\put(155.418, 28.528){\circle*{1}}
\put(155.618, 27.572){\circle*{1}}
\put(155.8, 26.613){\circle*{1}}
\put(155.95, 25.653){\circle*{1}}
\put(156.1, 24.694){\circle*{1}}
\put(156.25, 23.734){\circle*{1}}
\put(156.4, 22.771){\circle*{1}}
\put(156.75, 21.871){\circle*{1}}
\put(156.90, 20.971){\circle*{1}}
\end{picture}

\vskip 30pt
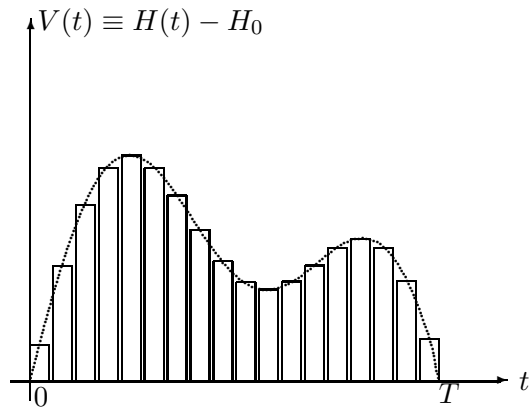
\noindent Fig. 2, A Hamiltonian variation and its pulse-like 
approximation.

\end{document}